\documentclass[preprint,12pt]{elsarticle}

\usepackage{amssymb}
\usepackage[usenames]{color}

\journal{Nucl.\ Instrum.\ Methods Phys.\ Res.\ A}

\begin{document}

\begin{frontmatter}



\title{Accessing interior magnetic field vector components in
neutron electric dipole moment experiments via exterior measurements \\
I.\ Boundary-value techniques}


\author{B.\ Plaster}
\address{Department of Physics and Astronomy, University of Kentucky, \\
Lexington, Kentucky 40506, USA}

\begin{abstract}
We propose a new concept for determining the interior magnetic field
vector components in neutron electric dipole moment experiments.  If a
closed three-dimensional boundary surface surrounding the fiducial
volume of an experiment can be defined such that its interior encloses
no currents or sources of magnetization, each of the interior vector
field components and the magnetic scalar potential will satisfy a
Laplace equation.  Therefore, if either the vector field components or
the normal derivative of the scalar potential can be measured on the
surface of this boundary, thus defining a Dirichlet or Neumann
boundary-value problem, respectively, the interior vector field
components or the scalar potential (and, thus, the field components
via the gradient of the potential) can be uniquely determined via
solution of the Laplace equation.  We discuss the applicability of
this technique to the determination of the interior magnetic field
components during the operating phase of neutron electric dipole
moment experiments when it is not, in general, feasible to perform
direct \textit{in situ} measurements of the interior field components.
We also study the specifications that a vector field probe must
satisfy in order to determine the interior vector field components to
a certain precision.  The technique we propose here may also be
applicable to experiments requiring monitoring of the vector magnetic
field components within some closed boundary surface, such as
searches for neutron-antineutron oscillations
along a flight path or measurements in storage rings
of the muon anomalous magnetic moment $g-2$ and the proton
electric dipole moment.
\end{abstract}



\end{frontmatter}

\section{Introduction}
\label{sec:intro}


The basic principle upon which all experimental searches for a neutron
electric dipole moment (EDM) employing stored ultracold neutrons (UCN)
are based concerns measurements of the neutrons' Larmor spin
precession frequencies $\nu_\pm$ in parallel ($+$) and anti-parallel
($-$) magnetic ($\vec{B}$) and electric ($\vec{E}$) fields,
\begin{equation}
h \nu_\pm = -2\left(\mu_n |\vec{B}| \pm d_n |\vec{E}|\right).
\end{equation}
Here, $\mu_n$ and $d_n$ denote the neutron's magnetic and electric
dipole moments, respectively.  A value for, or a limit
on, $d_n$ is then deduced from a comparison of the measured values of
$\nu_+$ and $\nu_-$.  The frequencies $\nu_+$ and $\nu_-$ are typically
determined either from sequential measurements in a single volume, or
from simultaneous measurements in separate volumes.  Therefore, a
central problem to all neutron EDM experiments concerns the
determination of the value of the magnetic field averaged over the
single or separate volumes, especially in the presence of temporal
fluctuations and/or spatial variations in the
field~\cite{lamoreaux09}.  An elegant solution providing for
real-time monitoring of the magnetic field is to deploy a so-called
``co-magnetometer'', whereby an atomic species with no EDM (or, at
least, one known to be significantly smaller than the neutron EDM)
co-habitates together with the stored UCN the fiducial volume
\cite{green98,baker13,golub94,altarev09,masuda12}.  The general idea
is then to carry out a measurement of the co-magnetometer atoms'
Larmor spin precession frequency in the magnetic field, from which the
temporal dependence of the \textit{scalar magnitude} of the magnetic
field $|\vec{B}|$ averaged over the fiducial volume is then deduced.

Thus, a co-magnetometer provides for a real-time, \textit{in situ}
measurement of the \textit{scalar magnitude} $|\vec{B}|$, which is
especially important for detecting any shifts in $|\vec{B}|$
correlated with the reversal of the direction of $\vec{E}$ relative to
$\vec{B}$.  However, there are many optimization parameters and
systematic effects in neutron EDM experiments associated with the
\textit{vector components} of the magnetic field, $B_i$, or,
equivalently, the field gradients $\partial B_i / \partial x_j$.  For
example, the longitudinal and transverse spin relaxation times, $T_1$
and $T_2$, the values of which contribute to a determination of an
experiment's statistical figure-of-merit, depend, among other
parameters, on the field gradients
\cite{gamblin65,schearer65,cates88a,cates88b,mcgregor90,golub11}.  As
another example, the dominant systematic uncertainty in the most
recent published limit on $d_n$ \cite{baker06} resulted from the
so-called ``geometric phase'' false EDMs of the neutron and the
co-magnetometer atoms
\cite{pendlebury04,harris06,lamoreaux05,barabanov06,golub08,yan11},
both of which are functions of the field gradients.

Despite the importance of knowledge of the field gradients in neutron
EDM experiments, the key point here is that a co-magnetometer does
not, in general, provide for a real-time, \textit{in situ} measurement
of the $\partial B_i / \partial x_j$ field gradients.  Nor is it
practical or feasible to carry out direct \textit{in situ}
measurements of the field components or field gradients in an
experiment's fiducial volume with some probe after the experimental
apparatus has been assembled.  However, the situation is not that
grim, as it has been shown that it may be possible to extract some
particular field gradients from measurements of the spin relaxation
times coupled with measurements of the neutrons' and co-magnetometer
atoms' trajectory correlation functions \cite{golub08}, and also
(under various assumptions on the symmetry properties of the magnetic
field profile) from a comparison of the neutron's and co-magnetometer
atoms' precession frequencies and their center-of-mass positions in
the magnetic field \cite{pendlebury04,masuda12}.

The concept we propose to employ for a real-time determination of the
interior vector field components $B_i$, and thus the field gradients
$\partial B_i / \partial x_j$, is a completely general method based on
boundary-value techniques which does not require any assumptions on
the symmetry properties (or lack thereof) of the field.  The basic
idea is to perform measurements of the field components on the surface
of a boundary surrounding the experiment's fiducial volume, and then
solve (uniquely) for the values of the field components in the region
interior to this boundary via standard numerical methods.
Although the physics basis of the concepts we discuss
in this paper are certainly not original (and likely known since the
origins of electromagnetic theory), to our knowledge this concept has
not been suggested for use in a neutron EDM experiment, although it
certainly has been suggested in other contexts (e.g., \cite{wind70});
nevertheless, we believe the discussion in this paper will be of value
to those engaged in neutron EDM experiments.  The remainder of this
paper is organized as follows.  In
Secs.\ \ref{sec:boundary_value_problem} and \ref{sec:discretization}
we discuss the boundary-value problem under consideration and its
applicability to neutron EDM experiments.  We then show examples from
numerical studies of this problem in Sec.\ \ref{sec:examples}
for the geometry of the neutron EDM experiment to be
conducted at the Spallation Neutron Source \cite{SNSEDM}, the concept
of which is based on the pioneering ideas of Golub and
Lamoreaux~\cite{golub94}.  We then study the specifications (e.g.,
precision) on a vector field probe in Sec.\ \ref{sec:specs}.  Finally,
we conclude with a brief summary in Sec.\ \ref{sec:summary}.

\section{Boundary-Value Problem for the Interior Vector Field Components}
\label{sec:boundary_value_problem}


\subsection{Statement of the Boundary-Value Problem}
\label{sec:boundary_value_problem_statement}

\begin{figure}
\begin{center}
\includegraphics[scale=0.50]{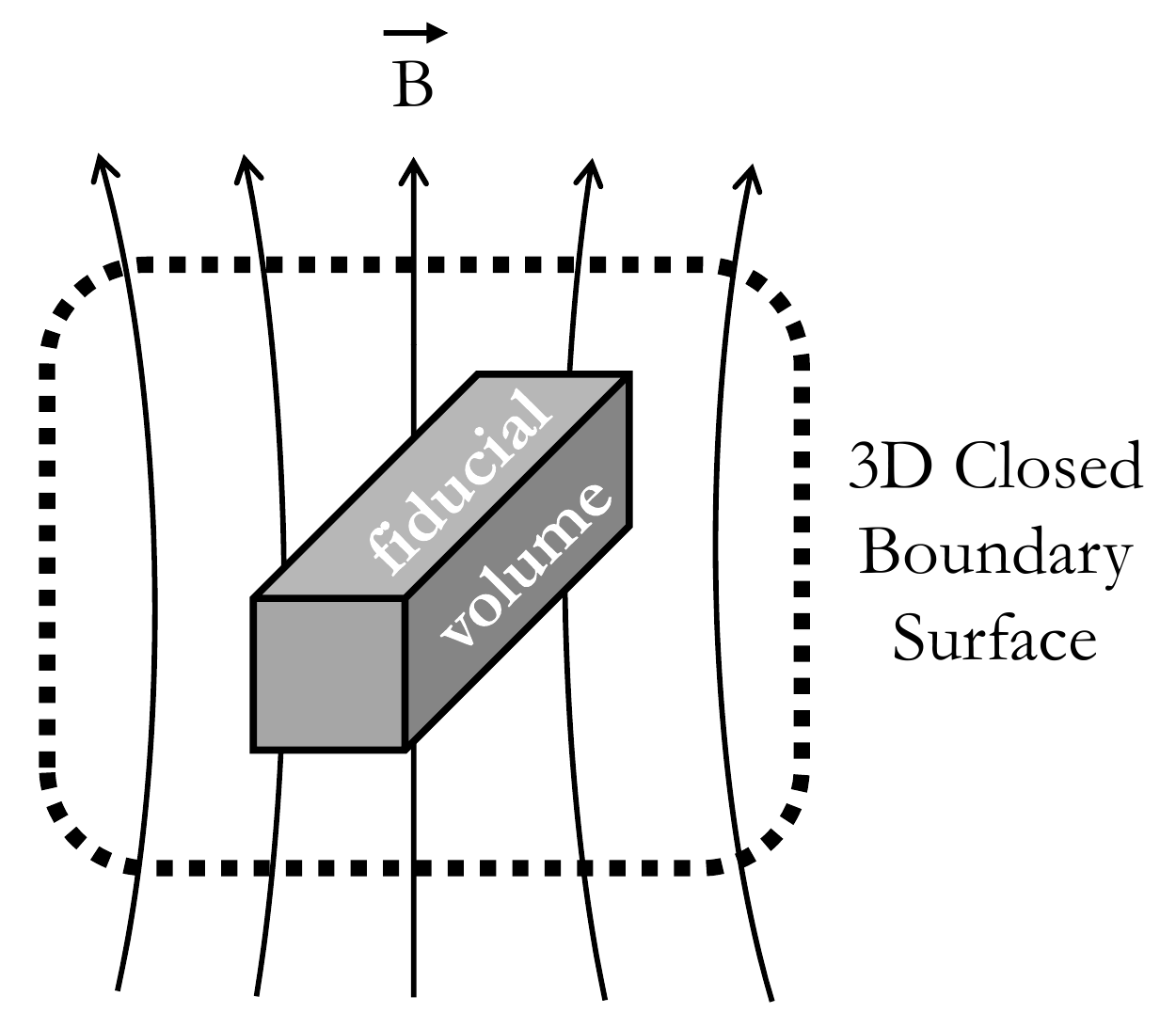}
\caption{Schematic illustration of the proposed boundary-value problem
for the determination of the magnetic field vector components interior
to a three-dimensional closed surface surrounding an experiment's
fiducial volume.}
\label{fig:boundary_value_problem_schematic}
\end{center}
\end{figure}

We begin by considering, as shown schematically in
Fig.\ \ref{fig:boundary_value_problem_schematic}, a closed
three-dimensional boundary surface surrounding the fiducial volume of
an experiment, which is situated within an arbitrary magnetic field
(i.e., no assumptions on the symmetry properties of the field are
necessary).  Our starting point is the fundamental equations of
magnetostatics, which in SI units are $\vec{\nabla} \times \vec{H} =
\vec{J}$ and $\vec{\nabla} \cdot \vec{B} = 0$, where $\vec{B} =
\mu_0(\vec{H} + \vec{M})$.  If we assume that the volume enclosed by
the boundary surface contains: (1) no sources of currents, such that
that the current density $\vec{J} = 0$ everywhere inside of the
boundary; and (2) no sources of magnetization, such that the
magnetization $\vec{M} = 0$ everywhere inside of the boundary, it then
follows that $\vec{\nabla} \times \vec{B} = 0$.  From this, we
immediately see, via application of the general vector identity
$\vec{\nabla} \times (\vec{\nabla} \times \vec{B}) =
\vec{\nabla}(\vec{\nabla} \cdot \vec{B}) - \vec{\nabla}^2 \vec{B}$,
that the magnetic field $\vec{B}$ (and, thus, each of its components
$B_i$) satisfies a Laplace equation,
\begin{equation}
\vec{\nabla}^2 \vec{B} = 0~~\Longrightarrow~~\vec{\nabla}^2 B_i = 0~~
(i=x,y,z),
\end{equation}
everywhere inside of the boundary.\footnote{Note that the latter
equality $\vec{\nabla}^2 B_i = 0$ is valid only if $\vec{B}$ is
expressed in terms of $(\hat{x},\hat{y},\hat{z})$ Cartesian
components.  This equality does not hold in curvilinear coordinates.
Therefore, we will use Cartesian coordinates exclusively hereafter.}

Alternatively, under the above assumptions that $\vec{\nabla} \times
\vec{B} = 0$ and $\vec{\nabla} \cdot \vec{B} = 0$ everywhere inside of
the boundary, in a manner analogous to charge-free electrostatics
(i.e., $\vec{\nabla} \times \vec{E} = 0$ and $\vec{\nabla} \cdot
\vec{E} = 0$) we can define a magnetic scalar potential $\Phi_M$ which
satisfies $\vec{B} = -\vec{\nabla}\Phi_M$.  From this, it then
immediately follows that imposing the requirement $\vec{\nabla} \cdot
\vec{B} = 0$ leads to a Laplace equation for the scalar potential,
\begin{equation}
\vec{\nabla}^2 \Phi_M = 0,
\end{equation}
everywhere inside of the boundary.

Therefore, in summary, we see that each of the vector field components
$B_i$ and the scalar potential $\Phi_M$ satisfy a Laplace equation
everywhere inside of the boundary, provided the boundary encloses no
current or magnetization.  Solutions to the Laplace equation, subject
to boundary values, are well known (e.g., \cite{jacksonEM}); thus,
determination of the interior field components or the scalar potential
from exterior boundary-value measurements is a solvable problem.

\subsection{Dirichlet Problem for the Interior Vector Components}
\label{sec:boundary_value_problem_dirichlet}

We now consider the Laplace equation for one of the vector components,
$\vec{\nabla}^2 B_i = 0$.  If boundary values for $B_i$ are known
everywhere on the surface of the boundary, the interior values of
$B_i$ everywhere inside the surface of the boundary can, in principle,
be obtained from an integral equation over the boundary values and the
appropriate Dirichlet Green's function for the geometry in question.
Thus, for the continuous version of the Dirichlet boundary-value
problem posed here, it is theoretically possible to solve for the
interior vector components everywhere inside the boundary, provided
their boundary values are known everywhere on the surface.  Such a
solution will be unique \cite{jacksonEM}.  Note that a limitation of
the Dirichlet problem we have formulated is that it requires boundary
values for the same component $B_i$ everywhere on the surface, with
the solution to the problem only yielding interior values for $B_i$
(i.e., no information on $B_j$ where $j \neq i$ can be deduced).

\subsection{Neumann Problem for the Interior Magnetic Scalar Potential}
\label{sec:boundary_value_problem_neumann}

Next we consider the Laplace equation for the magnetic scalar
potential, $\vec{\nabla}^2 \Phi_M = 0$.  The scalar potential $\Phi_M$
is, of course, not a physical observable; however, the vector
components of the gradient, $\vec{B} = -\vec{\nabla} \Phi_M$, are, of
course, physical observables.  Let $\hat{n}$ denote a unit vector
normal to the surface of the boundary.  If we then assume that
boundary values for the normal derivative of the scalar potential,
$\partial \Phi_M / \partial n = \vec{\nabla} \Phi_M \cdot \hat{n}$,
or, equivalently, the negative of the normal component of the magnetic
field, $-B_n = \partial \Phi_M / \partial n$, are known everywhere on
the surface of the boundary, the interior values of $\Phi_M$ can, in
principle, be obtained from an integral equation over the boundary
values and the appropriate Neumann Green's function for the geometry
in question.  Thus, for the continous version of the Neumann
boundary-value problem posed here, it is theoretically possible to
solve for the interior scalar potential everywhere inside the
boundary, provided the normal components of the magnetic field are
known everywhere on the surface.  Unlike the Dirichlet problem, the
solution to the Neumann problem for the interior scalar potential will
not be unique, as the value of the scalar potential is arbitrary up to
a constant $\Phi_M \rightarrow \Phi_M + \lambda$ \cite{jacksonEM};
however, the resulting interior magnetic field components, $\vec{B} =
-\vec{\nabla} \Phi_M$, will be unique.  Note that in contrast to the
Dirichlet problem, the solution to the Neumann problem determines
all of the interior vector components of $\vec{B}$.

\subsection{Comment on Exterior Measurements of $|\vec{B}|$}
\label{sec:boundary_value_problem_comment}

Exterior measurements (i.e., outside the fiducial volume) of the
scalar magnitude of the magnetic field, $|\vec{B}|$, are certainly
useful as they provide for important monitoring of the magnetic field
in the vicinity of the fiducial volume.  However, we note that such
measurements do not provide for a rigorous determination of either the
interior scalar magnitude $|\vec{B}|$ or the interior vector
components of $\vec{B}$, as the scalar magnitude $|\vec{B}|$ does not
satisfy a Laplace equation.  Therefore, any attempt to extract
information on the interior $\partial B_i / \partial x_j$ field
gradients from exterior measurements of $|\vec{B}|$ will necessarily
require various assumptions to be made on the symmetry properties of
the magnetic field.  In particular, fitting exterior measurements of
$|\vec{B}|$ to a multipole expansion in spherical harmonics in order
to determine interior values of $|\vec{B}|$ is not completely
rigorous, as such a multipole expansion is the solution for a quantity
which necessarily obeys the Laplace equation.

\section{Discretization of the Boundary-Value Problem}
\label{sec:discretization}

\subsection{Discretization of the Geometry}
\label{sec:discretization_geometry}


In the (hypothetical) continuous versions of the Dirichlet and Neumann
boundary-value problems formulated above, it was assumed that the
boundary values were known everywhere on the surface; this leads to
the well-known analytic solutions for the interior values in terms of
integral equations of Green's functions.  Of course, such a problem
cannot be realized in practice, as the boundary values can only be
determined at discrete measurement points.  Fortunately, numerical
solutions to discretized versions of the Dirichlet and Neumann
boundary-value problems are well known (e.g., \cite{numerical}).

\begin{figure}
\begin{center}
\includegraphics[scale=0.50]{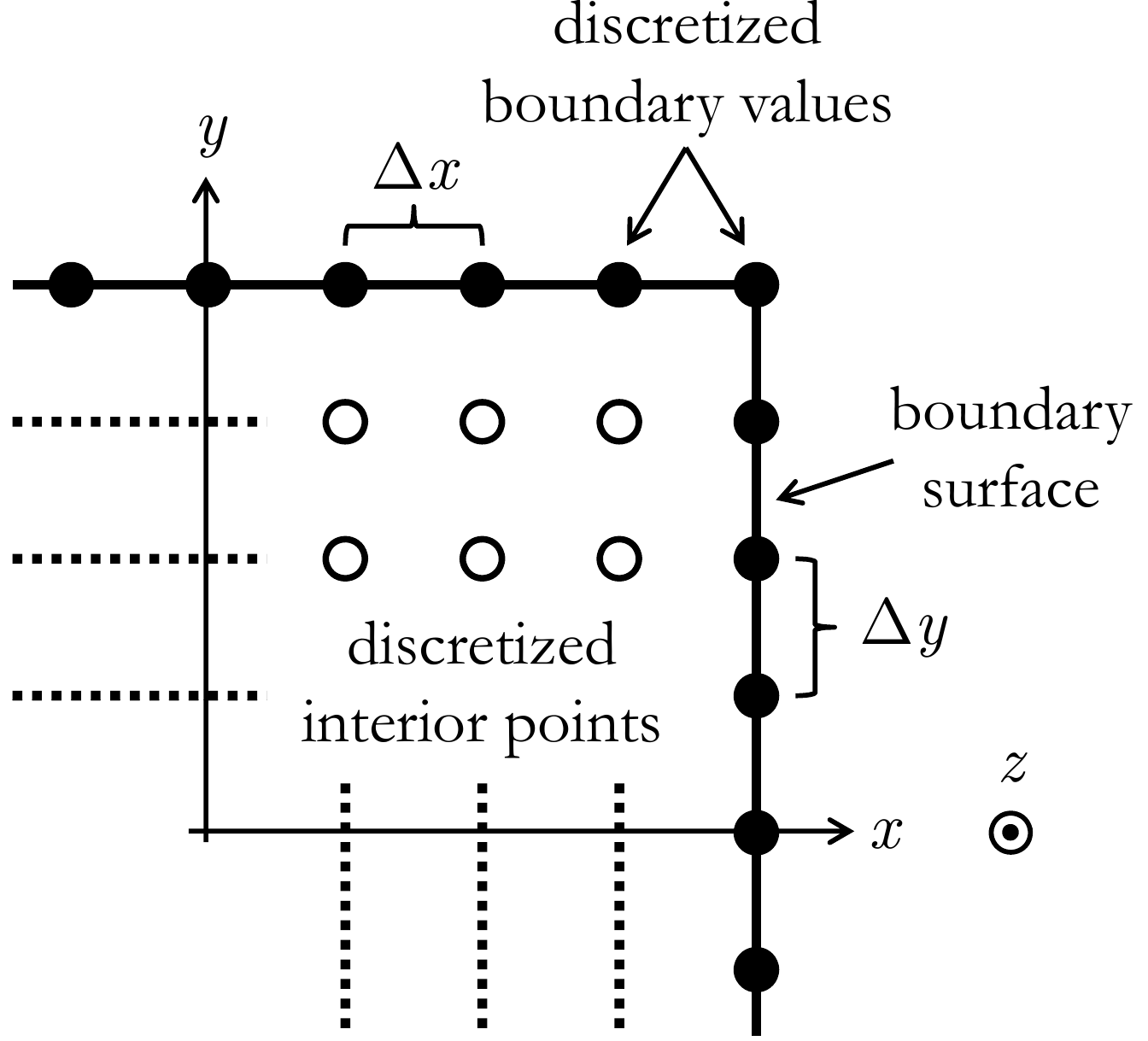}
\caption{Schematic illustration in two dimensions of the discretized
version of the boundary-value problem; the extension to three
dimensions is obvious.  Space is discretized into grid points, with
constant separations $\Delta x$, $\Delta y$, and $\Delta z$ along
their respective directions.  The boundary values are assumed to be
known over a grid of points on the surface of the boundary (filled
circles).  The solution is then desired over the grid of interior
points (open circles).}
\label{fig:boundary_value_problem_discretized}
\end{center}
\end{figure}

In the discretized versions of the boundary-value problems we will
consider hereafter, we will assume, as indicated schematically in
Fig.\ \ref{fig:boundary_value_problem_discretized}, that the boundary
values (i.e., $B_i$ for the Dirichlet problem or $B_n$ for the Neumann
problem) are known over a regularly-spaced grid on the surface of the
boundary, with the (constant) spacing between adjacent points along
the $x$, $y$, and $z$ directions denoted $\Delta x$, $\Delta y$, and
$\Delta z$.  Note that it is not necessary to employ uniform $\Delta x
= \Delta y = \Delta z$ grid spacings.  Also, it is not necessary to
employ ``flat'' boundary surfaces, such as the sides of a rectangular
box, although, for simplicity, the illustrative examples we will
consider in the next section do utilize a rectangular box geometry.
For example, one could discretize the surface of a torus, which would
be a natural candidate for a boundary surface surrounding the interior
of an experiment located within a circular accelerator storage ring.

Finally, it is also worthwhile to note that the boundary-value problem
must be cast in three dimensions.  For example, the solution to the
Laplace equation $\vec{\nabla}^2 B_i = 0$ need not satisfy 
$(\frac{\partial^2}{\partial x^2} +
\frac{\partial^2}{\partial y^2}) B_i = 0$
in two dimensions.  Therefore, an attempt to simplify
the Dirichlet and Neumann boundary-value problems for $B_i$ and
$\Phi_M$, respectively, from three to two dimensions will not, in
general, yield a valid solution.

\subsection{Methods for Numerical Solution of the Laplace Equation}
\label{sec:discretization_numerical_methods}

In general, there exists a multitude of techniques
for the numerical solution of the Laplace equation subject to boundary
values (see, e.g., \cite{numerical}), and we do not endeavor to
discuss these techniques here.  We employed the finite differencing
method of relaxation (examples of techniques include Jacobi iteration,
Gauss-Seidel iteration, successive overrelaxation, etc.), with the
results in the next section obtained using approximations to the
second-order partial derivatives valid to $\mathcal{O}((\Delta x)^2)$,
i.e.,
\begin{equation}
\left.\frac{\partial^2 u}{\partial x^2}\right|_{(i,j,k)} =
\frac{u(i+1,j,k) - 2u(i,j,k) + u(i-1,j,k)}{(\Delta x)^2}.
\end{equation}
Here the notation $u(i,j,k)$ denotes the solution to the Laplace
equation $\vec{\nabla}^2 u(x,y,z) = 0$ at some $(x,y,z)$ grid point
indexed by the integers $(i,j,k)$.  Note that to this order, if one
takes $\Delta x = \Delta y = \Delta z$, one obtains the well-known
result for $u(i,j,k)$ in terms of the values of the solution at its
six nearest neighbor grid points,
\begin{eqnarray}
u(i,j,k) &=& \frac{1}{6}\Big[
u(i+1,j,k) + u(i-1,j,k) + u(i,j+1,k) + u(i,j-1,k) \nonumber \\
&&~~~+ u(i,j,k+1) + u(i,j,k-1) \Big].
\label{eq:laplace_grid_solution}
\end{eqnarray}

\section{Examples from Numerical Studies}
\label{sec:examples}


\subsection{Geometry and Magnetic Field}
\label{sec:examples_geometry}

As a validation of our concept, we now show results from numerical
studies of the Dirichlet boundary-value problem for $B_i$ and the
Neumann boundary-value problem for $\Phi_M$.  The
example geometry we will consider is that of the neutron EDM
experiment to be conducted at the Spallation Neutron Source
\cite{SNSEDM}.  In particular, this geometry consists of two
rectangular measurement volumes, which together span our definition of
a rectangular fiducial volume of dimensions 25 cm ($-12.5$ cm $<x<$
$12.5$ cm) $\times$ 10 cm ($-5.0$ cm $<y<$ $5.0$ cm) $\times$ 40 cm
($-20.0$ cm $<z<$ $20.0$ cm).  We then employ a rectangular boundary
surface of dimensions 80 cm ($-40.0$ cm $<x<$ $40.0$ cm) $\times$ 80
cm ($-40.0$ cm $<y<$ $40.0$ cm) $\times$ 100 cm ($-50.0$ cm $<z<$
$50.0$ cm).  Thus, the volume enclosed by the boundary surfaces is
significantly larger (factor of 64) than the fiducial volume, with the
boundary surfaces all located $\sim 30$ cm from the fiducial volume.

The magnetic field we will consider is a calculated field map of a
modified $\cos\theta$ coil\footnote{Note that the field map we
employed was calculated for this work by M.\ P.\ Mendenhall for the
geometry parameters of the modified $\cos\theta$ coil described in
\cite{perezgalvan11}, but without its surrounding
cylindrically-concentric ferromagnetic shield, as such a calculation
would have required significantly more computing time.  We chose to
use this field for our example because the field shape is not trivial;
as can be seen later, the field shape is quartic near the origin.}
under development for this particular experiment
\cite{perezgalvan11}.  The orientation of the $\cos\theta$ coil is
such that the fiducial volume is centered on the coil's center, with
the magnetic field $\vec{B}$ oriented along the $x$-direction at the
center of the fiducial volume.

\subsection{Example Dirichlet Problem: Densely-Spaced Boundary Values}
\label{sec:examples_dirichlet_dense}

As our first numerical example, we considered a Dirichlet
boundary-value problem for each of the $(B_x,B_y,B_z)$ field
components in a geometry where the spacing between the grid points is
$\Delta x = \Delta y = \Delta z = 1.0$ cm, thus resulting in 44,802
densely-spaced grid points on the surface of the boundary.  As per the
discussion in
Sec.\ \ref{sec:discretization}, we assumed
the values of $(B_x,B_y,B_z)$ were known at all of the 44,802 boundary
grid points.  We then proceeded to solve for the values of
$(B_x,B_y,B_z)$ at all of the 617,859 interior grid points.  The
computing time required for $10^5$ iterations of our C++ code on a
Linux machine was 93 minutes.  Obviously, implementing such a
densely-spaced configuration would not be possible or practical in an
actual experiment; instead, the point of this hypothetical example was
to first demonstrate the validity of the boundary-value technique
for the determination of the interior field components.

\begin{figure}
\begin{center}
\includegraphics[scale=0.62]{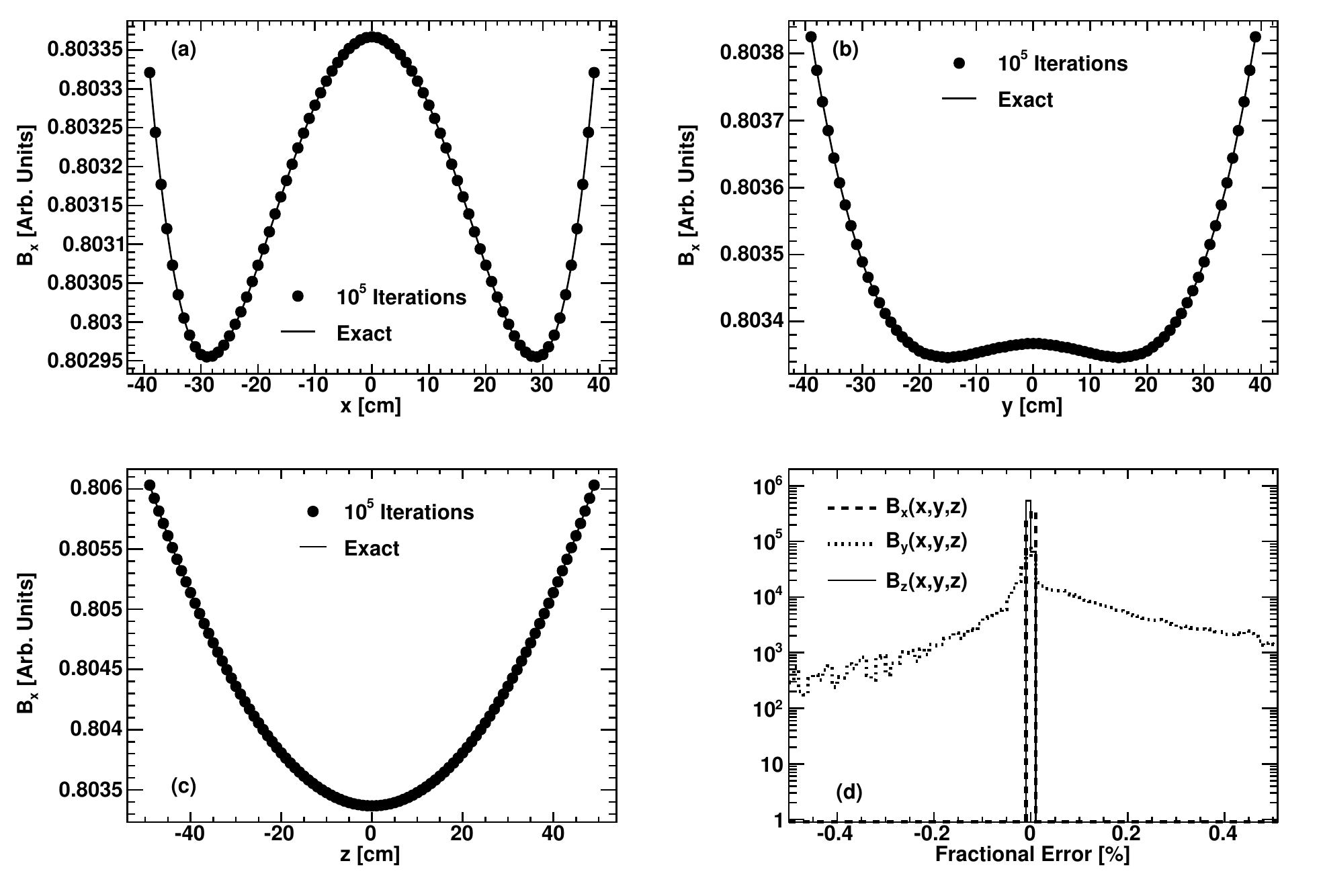}
\caption{Results from numerical studies of the Dirichlet boundary-value
problem for $B_i$ for the densely-spaced grid of $\Delta x =
\Delta y = \Delta z = 1.0$ cm (see text for details).  Calculated
interior values for $B_x$ along the $x$-, $y$-, and $z$-axes are shown
in panels (a), (b), and (c) as the filled circles, and are compared
with the exact values shown as the solid curves.  Panel (d) shows a
histogram of the fractional error in the calculated interior values of
$(B_x,B_y,B_z)$ for all of the interior grid points.}
\label{fig:case1_figures}
\end{center}
\end{figure}

\begin{figure}
\begin{center}
\includegraphics[scale=0.62]{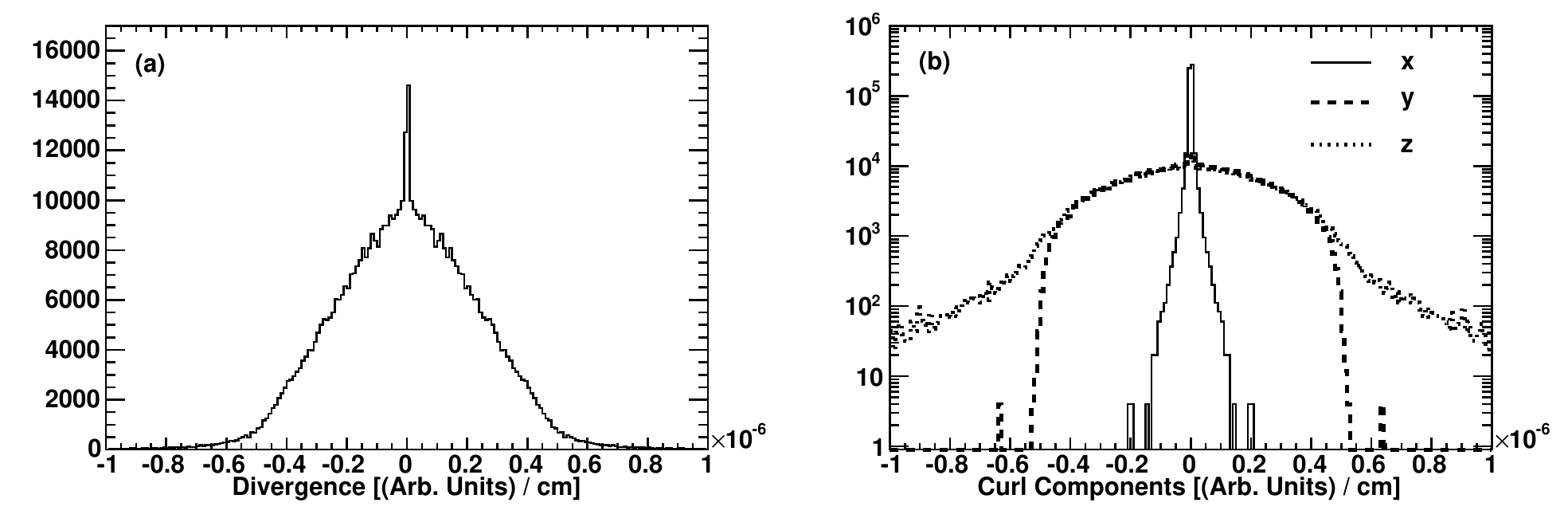}
\caption{Histograms of values for $\vec{\nabla} \cdot \vec{B}$ in
panel (a) and the $x$-, $y$-, and $z$-components of $\vec{\nabla}
\times \vec{B}$ in panel (b) for the calculated interior values of
$B_i$ at all of the interior grid points from the Dirichlet boundary-value
problem for the densely-spaced grid of $\Delta x = \Delta y =
\Delta z = 1.0$ cm (see text for details).}
\label{fig:case1_del}
\end{center}
\end{figure}

The results of this exercise are shown in
Fig.\ \ref{fig:case1_figures}.  Panels (a), (b), and (c) compare the
calculated interior values of $B_x$ along the $x$-, $y$-, and $z$-axes
with the exact values from the field map, and panel (d) then shows
histograms of the fractional errors [defined to be (calculated $-$
exact)/exact] in the calculated interior values of $(B_x,B_y,B_z)$ at
all of the interior points.  The agreement between the
calculated and exact values is seen to be excellent, thus clearly
demonstrating the validity of our proposed concept.  As a further
check, Fig.\ \ref{fig:case1_del} shows histograms of values for
$\vec{\nabla} \cdot \vec{B}$ and $\vec{\nabla} \times \vec{B}$
determined from the calculated interior values (using the centered
difference approximation).  As expected, the distributions are
centered on zero, consistent with the initial assumptions of the
problem.

\subsection{Example Neumann Problem: Densely-Spaced Boundary Values}
\label{sec:examples_neumann_dense}

\begin{figure}
\begin{center}
\includegraphics[scale=0.62]{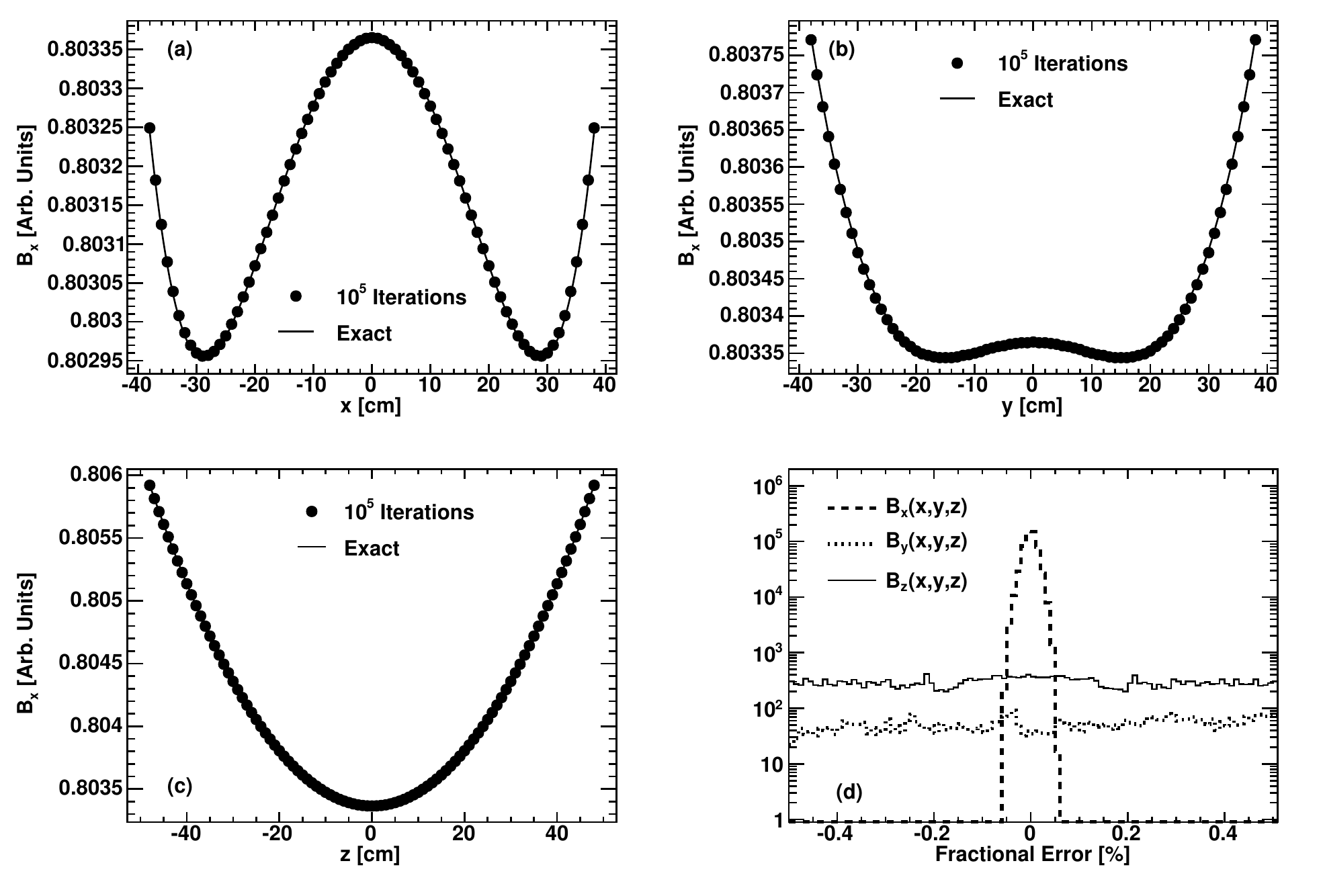}
\caption{Results from numerical studies of the Neumann boundary-value
problem for $\Phi_M$ and resulting values for $\vec{B} =
-\vec{\nabla}\Phi_M$ for the densely-spaced grid of $\Delta x = \Delta
y = \Delta z = 1.0$ cm (see text for details).  Values for $B_x =
-\partial \Phi_M / \partial x$ along the $x$-, $y$-, and $z$-axes as
calculated from the interior values for $\Phi_M$ are shown in panels
(a), (b), and (c) as the filled circles, and are compared with the
exact values shown as the solid curves.  Panel (d) shows a histogram
of the fractional error in the calculated interior values of
$(B_x,B_y,B_z)$ for all of the interior grid points.}
\label{fig:case5_figures}
\end{center}
\end{figure}

We now consider the Neumann boundary-value problem for $\Phi_M$ for
the same densely-spaced grid configuration employed in the discussion
of the Dirichlet boundary-value problem in Section
\ref{sec:examples_dirichlet_dense}.  Again, as per the discussion in
Sec.\ \ref{sec:discretization}, we assumed
the values of $-B_n = \partial \Phi_M / \partial n$ were known at all
of the boundary grid points.  We then proceeded to solve for
the values of $\Phi_M$ at all of the interior grid points.
The computing time required for $10^5$ iterations of our C++ code on a
Linux machine for the solution of the Neumann problem for $\Phi_M$ was
24 minutes, a little better than 1/3 of that required for solution of
the Dirichlet problem for all three components of $\vec{B}$.

The results of this exercise are shown in
Fig.\ \ref{fig:case5_figures}.  As before, panels (a), (b), and (c)
compare the calculated interior values of $B_x = -\partial \Phi_M /
\partial x$ with the exact values from the field map, and panel (d)
then shows histograms of the fractional errors in the calculated
interior values of $(B_x,B_y,B_z)$ for all of the interior grid
points.  Again, the agreement between the calculated and exact values
for $B_x$ (i.e., the dominant field component) is excellent, again
clearly demonstrating the validity of the Neumann concept.  However,
the fractional errors in the calculated values of $B_y$ and $B_z$ are
larger than those for $B_x$; this is the result of a loss of precision
in calculating these significantly smaller components via derivatives
of $\Phi_M$.

\subsection{Example Dirichlet Problem: Coarsely-Spaced Boundary Values}
\label{sec:examples_dirichlet_coarse}

\begin{figure}
\begin{center}
\includegraphics[scale=0.62]{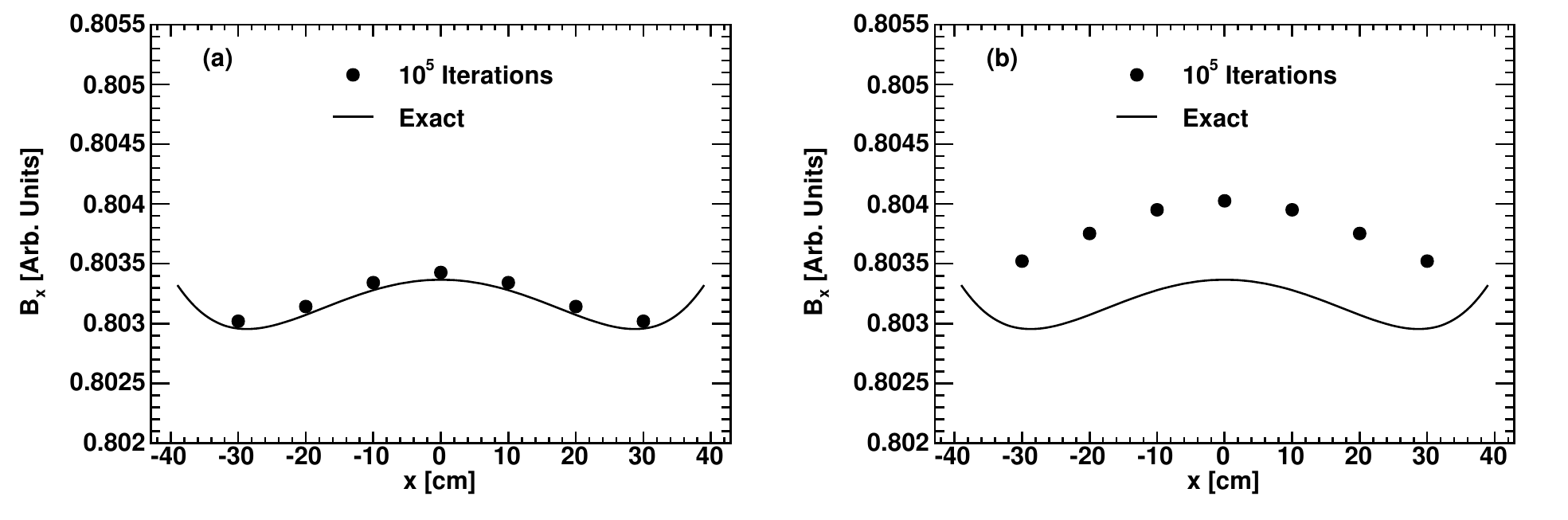}
\caption{Results from numerical studies of the Dirichlet
boundary-value problem for $B_i$ for two coarsely spaced boundary
value grids.  Panel (a) shows calculated interior values of $B_x$
along the $x$-axis (filled circles) compared with the exact values
(solid curves) for a grid with $(\Delta x, \Delta y, \Delta z)$ = (10
cm, 10 cm, 50 cm).  Panel (b) is for a grid with $(\Delta x, \Delta y,
\Delta z)$ = (10 cm, 40 cm, 50 cm).  Note that we do not show values
for $B_x$ along the $y$- or $z$-axes, as there are very few interior
grid points along these dimensions given the relatively large $\Delta
y$ and $\Delta z$ grid spacings.}
\label{fig:coarse_figures}
\end{center}
\end{figure}

We now consider more realistic examples of the Dirichlet
boundary-value problem in which the grids are (significantly) more
coarsely spaced than those of the previous examples.  First,
calculated interior values of $B_x$ along the $x$-axis are shown in
panel (a) of Fig.\ \ref{fig:coarse_figures} for a grid with spacings
$(\Delta x, \Delta y, \Delta z)$ = (10 cm, 10 cm, 50 cm), which
would require measurements of 194 boundary values.  The agreement
between the calculated and exact values is still quite good.  Second,
panel (b) shows results from the same calculation for an even coarser
grid with spacings
$(\Delta x, \Delta y, \Delta z)$ = (10 cm, 40 cm, 50 cm),
requiring measurements of 74 boundary values.  The
agreement is now somewhat degraded, although the calculated and exact
values still agree to the level of $\sim 0.08$\%.  A drawback of this
latter coarse grid is that the number of interior points are limited
to those shown in panel (b) because $\Delta y$ and $\Delta z$ are
simply half of the extent of the fiducial volume in their respective
directions.

The computing time required for $10^5$ iterations of our codes
was $< 10$ seconds for both of these coarse grids.

\section{Specifications on the Vector Field Probe}
\label{sec:specs}

Measurements of boundary values in an experiment with a vector field
probe will, of course, be subject to noise and/or systematic errors
such as uncertainties in the probe's $(x,y,z)$ positioning or its
calibration.  To study the specifications that a probe must satisfy in
order to determine the interior field components to a certain
precision, we employ a simple model in which we subject each boundary
value $B_i$ to a Gaussian fluctuation parameter $\delta$,
\begin{equation}
B_i \rightarrow B_i (1 + \delta),
\label{eq:gaussian_fluctuation}
\end{equation}
where $\delta$ is randomly sampled from a Gaussian with a mean of zero
and a particular width $\sigma$.  This simple model accounts for noise
fluctuations in the measurement of $B_i$ and also errors in the
probe's $(x,y,z)$ positioning, the latter of which can be interpreted
as equivalent to an error in the measurement at the nominal $(x,y,z)$
position.

\begin{figure}
\begin{center}
\includegraphics[scale=0.62]{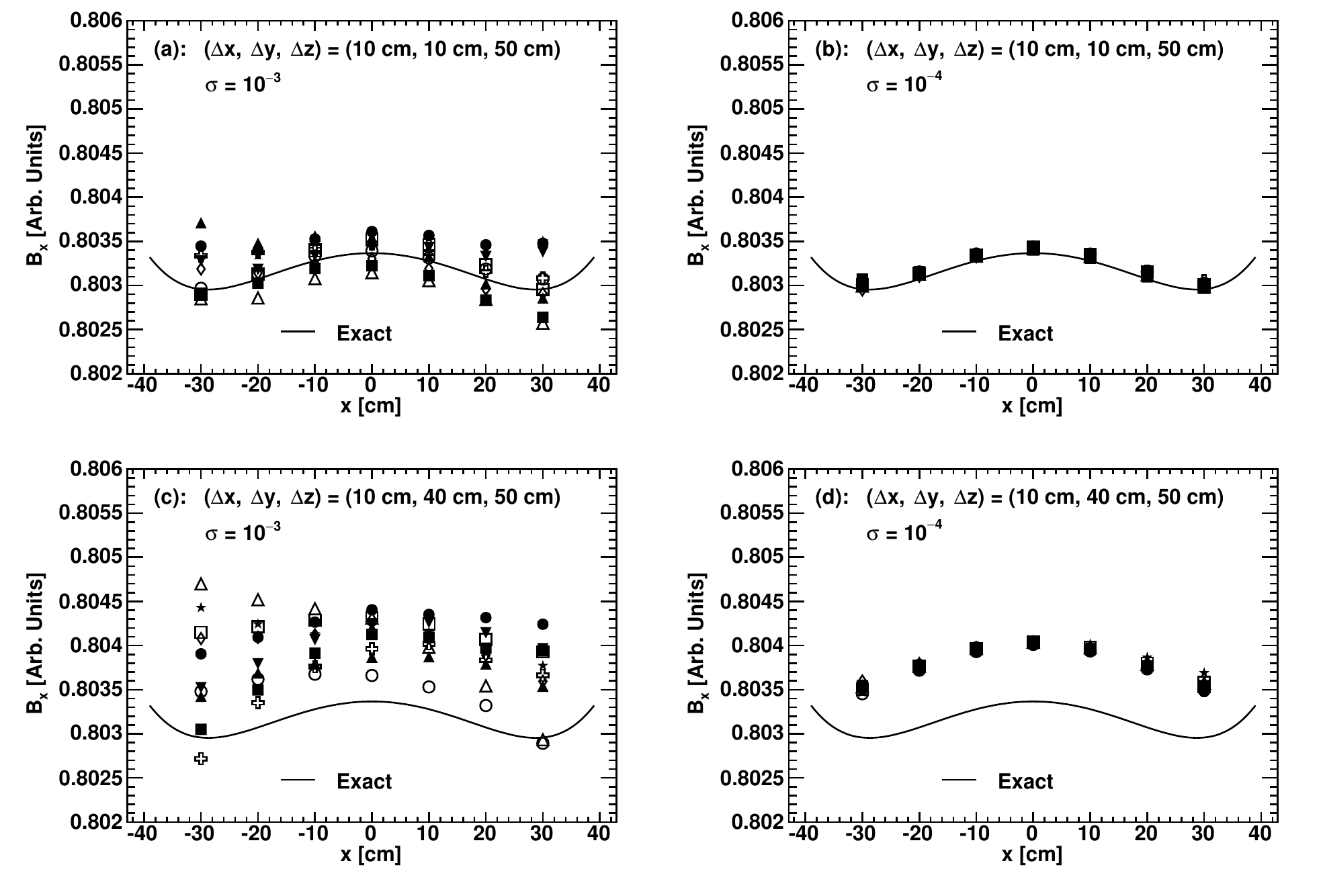}
\caption{Results from numerical studies of the impact of noise and/or
systematic errors in the measurements of the boundary values on the
calculated interior values.  All panels show calculated interior
values of $B_x$ along the $x$-axis for ten random configurations of
boundary values (indicated by the different data symbols) generated
according to the Gaussian fluctuation model discussed in the text.
Panel (a): grid spacing of $(\Delta x, \Delta y, \Delta z)$ = (10 cm,
10 cm, 50 cm) and Gaussian fluctuation parameter $\sigma = 10^{-3}$.
Panel (b): (10 cm, 10 cm, 50 cm) and $\sigma = 10^{-4}$.  Panel (c):
(10 cm, 40 cm, 50 cm) and $\sigma = 10^{-3}$.  Panel (d): (10 cm, 40
cm, 50 cm) and $\sigma = 10^{-4}$.  In panels (b) and (d) the
different data symbols all overlap each other.}
\label{fig:sigma_noise}
\end{center}
\end{figure}

We considered two examples of $\sigma = 10^{-3}$ and $10^{-4}$ which
we illustrate within the context of the two coarse
grids discussed previously in Section
\ref{sec:examples_dirichlet_coarse} (i.e., those with $(\Delta x,
\Delta y, \Delta z)$ = (10 cm, 10 cm, 50 cm) and (10 cm, 40 cm, 50
cm), yielding 194 and 74 boundary values, respectively).  To provide
context for an experiment, a $\sigma$ of $10^{-4}$ would correspond to
a Gaussian width of $10^{-6}$ Gauss on a $10^{-2}$ Gauss field value,
where $10^{-2}$ Gauss is the typical scale of field magnitudes in
recent and future neutron EDM experiments.  For each of these
$\sigma$ values, we generated ten random configurations of boundary
values in which each of the boundary values was subjected to a
Gaussian fluctuation according to
Eq.\ (\ref{eq:gaussian_fluctuation}).  The impact of these
fluctuations on the calculated interior values is shown in
Fig.\ \ref{fig:sigma_noise}, where we show the calculated interior
values of $B_x$ along the $x$-axis for each of the ten random
configurations.  As can be seen there, if $\sigma = 10^{-3}$ the
spread in the calculated interior values is rather large (and the sign
of the gradient $\partial B_x / \partial x$ that would be deduced
would be incorrect in some cases), whereas if $\sigma = 10^{-4}$ the
spread is small and any differences in the values of $\partial B_x /
\partial x$ deduced from the calculated interior values would be
small.

Thus, within the context of this simple model, we conclude that a
reasonable specification on a vector field probe is that the relative
uncertainties in the probes' measurements of the boundary values must
be of order $10^{-4}$ and any errors in the $(x,y,z)$ positioning of
the probes must not result in measured field values that differ by
more than $10^{-4}$ from what their values would be at their nominal
positions.

\section{Summary}
\label{sec:summary}


In summary, we have proposed a new concept for determining the
interior magnetic field vector components in neutron EDM experiments
via Dirichlet and Neumann boundary-value techniques, whereby exterior
measurements of the field components over a closed boundary surface
surrounding the experiment's fiducial volume uniquely determine the
interior field components via solution of the Laplace equation.  We
suggest that this technique will be of particular use to neutron
EDM experiments after they have been assembled and are in operation,
when it is no longer possible to perform an in-situ field map.

We also emphasize that this technique is certainly not limited in its
applicability to neutron EDM experiments.  Indeed, this technique
could be of interest of any experiment requiring monitoring of vector
field components within some well defined boundary surface.  Some
examples of this could be experimental searches for
neutron-antineutron ($n\overline{n}$) oscillations along a flight
path or experiments utilizing storage rings for measurements of the
muon $g-2$ or the proton EDM.  The concept for an
$n\overline{n}$ experiment would be to mount field probes along the
neutron flight path in the region interior to the magnetic shielding,
and for the storage ring experiments on the beam vacuum pipe in the
region interior to the storage ring magnets and electrodes.

However, as relevant for neutron EDM experiments, we do note that one
limitation of our boundary-value concept was discussed in
Sec.\ \ref{sec:examples_dirichlet_coarse}: that is, the number of
interior points at which the interior fields can be calculated (and,
thus, the resolution at which the field gradients can be determined)
is limited by the number of grid points (or, equivalently, the grid
spacing) at which the boundary values are measured.  In a forthcoming
work \cite{nouri_plaster}, we will explore an alternative
technique of fitting measurements of exterior field components to a
multipole expansion of the field components or the magnetic scalar
potential.  Such a technique is valid because the field components and
the scalar potential satisfy the Laplace equation, and an expansion in
multipoles is a valid solution to the Laplace equation.  This
technique, via the nature of a ``fit'' (as compared to the direct
solution of the Laplace equation in the boundary-value technique
discussed in the present work), holds the potential for a
determination of the interior field components everywhere within the
fiducial volume. \\

\newpage
\noindent\textbf{Acknowledgments} \\
\indent We thank M.\ P.\ Mendenhall for providing the field map of the
$\cos\theta$ coil we used in our example calculations.  We thank
C.~Crawford, B.~Filippone, R.~Golub, and J.~Miller for several
valuable suggestions regarding the development of the concept,
R.\ W.\ Pattie,
Jr.\ for suggesting the possible applicability of the concept
to $n\overline{n}$ experiments, and R.\ Golub, M.\ E.\ Hayden,
S.\ K.\ Lamoreaux, and N.\ Nouri
for comments on the manuscript.  This work was supported in
part by the U.\ S.\ Department of Energy Office of Nuclear Physics
under Award No.\ DE-FG02-08ER41557.



\end{document}